\documentclass[aps,showpacs,english,preprint]{revtex4}

\usepackage{mathrsfs}
\usepackage{graphicx}
\usepackage{amsmath, amssymb}
\usepackage{babel}

\begin{document}

\title{Tunnelling between edge states of a 2D topological insulator and a Fermi
liquid lead through a quantum dot}

\author{Chien-Yeah Seng}

\author{Tai-Kai Ng}

\affiliation{Department of Physcis, Hong Kong University of Science
and Technology, Hong Kong, People's Republic of China}

\date{6 August 2011}

\begin{abstract}
 In this paper we study a non-equilibrium resonant tunnelling problem where a non-interacting quantum dot is connected to
 two leads, one being the edge of an interacting 2-D topological insulator (Luttinger liquid) and the other being a usual
 Fermi liquid. We show that the current passing through the system can be expressed in terms of a non-equilibrium local
 single-particle Green's function of the Luttinger liquid lead which can be analysed using standard bosonization-Renormalization
 Group (RG) technique. In particular, some exact results can be extracted in the small bias limit with repulsive
 electron-electron interaction. A simple formula which captures the qualitative feature of the I-V relation over whole
 temperature and voltage bias range is being proposed and studied.
\end{abstract}


\maketitle


  The discovery of topological insulators\cite{topological insulator1,topological insulator2} has generated much interests
  and activities in the condensed matter physics community. In the case of 2-D topological insulators (Quantum Spin-Hall
  systems), helical edge states respecting time-reversal symmetry\cite{SCZhang} exist and provide an example of interacting
  Luttinger liquid which can be studied experimentally up to room temperature.
  The transport properties of interacting one-dimensional systems have been studied extensively in the
  literature\cite{Kane&Fisher,Furusaki&Nagaosa,Fabrizio&Gogolin}, especially after the discovery of edge states in fractional
  quantum hall (FQH) liquids \cite{Res-Tun in FQH} (chiral Luttinger liquids). For
  example, the problem of non-interacting quantum dot connected to leads of FQH edge states was studied by Chamon and
  Wen\cite{Chamon & Wen}, and later by Furusaki \cite{Furusaki} at temperatures higher than the tunnelling strength.

\begin{figure}
\includegraphics[scale=0.45]{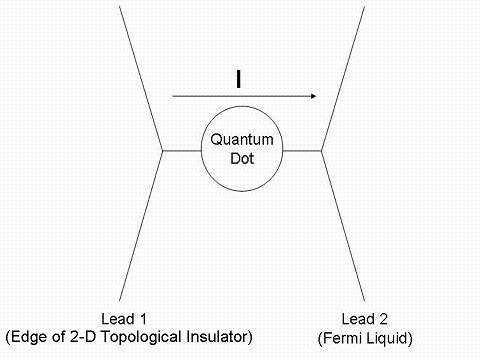}\caption{\label{fig:The-Configuration}The configuration}
\end{figure}

  The emergence of topological insulator makes these 1-D theoretical models realizable. In this
  paper we study a simple model of a non-interacting quantum dot connected to two leads, one being the edge of 2-D
  topological insulator with electron-electron interaction, the other being a normal Fermi liquid lead
  (Fig.\ref{fig:The-Configuration}). We show here that the non-linear I-V relation of the system can be expressed in terms of a
  non-equilibrium local single-particle Green's function of the topological insulator lead which can be
  analyzed in the limit of small voltage bias and low temperature using a conventional renormalization-group (RG) approach.
  A few exact results are extracted from the RG analysis and a simple formula which captures the qualitative feature
  of the I-V relation over whole temperature and voltage bias range is being proposed and studied in this paper.

 We start with the model Hamiltonian
 $H=H_{01}+H_{02}+H_{0d}+H_{T}$, where
 \begin{subequations}
 \begin{eqnarray}
H_{01} & = & -iv_{F}\int dx(\hat{\psi}_{1R,\uparrow}^{+}(x)\partial_{x}\hat{\psi}_{1R,\uparrow}(x)\nonumber\\
& & -\hat{\psi}_{1L,\downarrow}^{+}(x)\partial_{x}\hat{\psi}_{1L,\downarrow}(x))\nonumber\\
 &  & +\frac{g}{2}\int
 dx(\hat{\psi}_{1R,\uparrow}^{+}(x)\hat{\psi}_{1R,\uparrow}(x)\nonumber\\
&&+\hat{\psi}_{1L,\downarrow}^{+}(x)\hat{\psi}_{1L,\downarrow}(x))^{2}
\end{eqnarray}
and
\begin{equation}
 H_{02}=\sum_{k,\sigma}\varepsilon_{2k}\hat{c}_{2k,\sigma}^{+}\hat{c}_{2k,\sigma}
 \end{equation}
 are terms in the Hamiltonian describing the edge of a 2-D topological insulator ($H_{01}$) and the Fermi
 liquid lead ($H_{02}$), respectively.
 \begin{equation}
 H_{0d}=\varepsilon_{0}\sum_{\sigma}\hat{d}_{\sigma}^{+}\hat{d}_{\sigma}
\end{equation}
 is the Hamiltonian describing the quantum dot and
\begin{eqnarray}
H_{T} & = & T_{1}\{\hat{\psi}_{1R,\uparrow}^{+}(0)\hat{d}_{\uparrow}+\hat{\psi}_{1L,\downarrow}^{+}(0)\hat{d}_{\downarrow}+h.c\}\nonumber\\
 &  &
 +T_{2}\{\hat{\psi}_{2,\uparrow}^{+}(0)\hat{d}_{\uparrow}+\hat{\psi}_{2,\downarrow}^{+}(0)\hat{d}_{\downarrow}+h.c\}
 \label{eq:Hamiltonian}
\end{eqnarray}
\end{subequations}
 describes electron tunnelling between the dot and the leads. We note that since the propagation and spin directions of
 electrons in the topological insulator lead
 are tied together (helical edge), we can forget its spin suffix and the helical edges states of topological insulator
 behave like a spinless Luttinger Liquid \cite{Luttinger Liquid,vic}.


 $H_{01}$ can be transformed to a free boson system by standard bosonization technique \cite{Giamarchi} where the fermion
 correlation function $g_{1}(t,t')\equiv-i\left\langle T_{C}\hat{\psi}_{1,\sigma}(0,t)\hat{\psi}^{+}_{1,\sigma}(0,t')\right
 \rangle_{H_{01}}$ can be computed straightforwardly\cite{My thesis}. In bosonization theory a Luttinger liquid is
 usually characterized by the interaction strength parameter $K\equiv (1+\frac{g}{\pi v_{F}})^{-1/2}$; here we
 further define $\vartheta\equiv \frac{1}{2}(K+\frac{1}{K}-2)$ which is a parameter we shall use frequently in the following.

 The DC current $I$ flowing from lead 1 to 2 in the above system can be expressed as\cite{Meir&Wingreen&Jauho}
\begin{eqnarray}
I &=&-I_{2}=e\sum_{\sigma}\left\langle\frac{dN_{2\sigma}}{dt}\right\rangle\nonumber\\
& = & -ieT_{2}\sum_{\sigma}\left\langle
\hat{\psi}_{2,\sigma}^{+}(0,t)\hat{d}_{\sigma}(t)-\hat{d}^{+}_{\sigma}(t)\hat{\psi}_{2,\sigma}(0,t)\right\rangle
\label{eq:starting form of current}\end{eqnarray}
 where $I_2$ is the current flowing from lead 2 to the quantum dot. Since lead 2 is non-interacting, we can eliminate the
 lead-2 electron operators through their equation of motion. Following Ref.\cite{Meir&Wingreen&Jauho} we obtain after some
 straightforward algebra
\begin{eqnarray} I&=&-e|T_{2}|^{2}\sum_{\sigma}\int
d\varepsilon\rho_{2}(\varepsilon)[2f(\varepsilon-\mu_{2})\mathrm{Im}\{G_{dd,\sigma}^{R}(\varepsilon)\}\nonumber\\
&&+\mathrm{Im}\{G_{dd,\sigma}^{<}(\varepsilon)\}]
\label{eq:generalized Wingreen formula}\end{eqnarray}
 where $G_{dd,\sigma}^{R}(\varepsilon)$ and $G_{dd,\sigma}^{<}(\varepsilon)$ are the Fourier
 Transform of the ``retarded'' and ``less'' components of the on-site (Keldysh) Green's
 function of the quantum dot, $G_{dd,\sigma}(t,t')\equiv-i\left\langle T_{C}\hat{d}_{\sigma}(t)\hat{d}^{+}_{\sigma}
 (t')\right\rangle$, where $t$ and $t'$ are time-ordered along a closed time contour from $t=-\infty$ to
 $t=\infty$ and then back to $t=-\infty$. $\rho_{2}(\varepsilon)$ and $\mu_2$ are the density of states (which will be taken
 to be a constant later) and chemical potential of lead 2, respectively. $f(\varepsilon)$ is the Fermi-distribution function.
 Notice that the current is expressed solely in terms of a (non-equilibrium) local single-particle Green's function of the
 system. This result is possible because lead 2 is non-interacting.

 To calculate $G_{dd,\sigma}$ we follow the Keldysh path integral formalism \cite{Kamenev}.
 The action of the system is \cite{Vurkevich}\begin{eqnarray}
 S&=&\int_{C}dt[\int dx\sum_{\sigma} \{\bar{\psi}_{1,\sigma}(x,t)i(\partial_{t}+\sigma
 v_{F}\partial_{x}) \psi_{1,\sigma}(x,t)\}\nonumber\\
&&-\frac{g}{2}\int dx(\sum_{\sigma}\bar{\psi}_{1,\sigma}(x,t)\psi_{1,\sigma}(x,t))^{2}\nonumber\\
&&+\sum_{k,\sigma}\bar{c}_{2k,\sigma}(t)(i\partial_{t}-\varepsilon_{2k})c_{2k,\sigma}(t)\nonumber\\
&&+\sum_{\sigma}\{T_{1}\bar{\psi}_{1,\sigma}(0,t)d_{\sigma}(t)+T_{2}\bar{\psi}_{2,\sigma}(0,t)d_{\sigma}(t)+c.c\}\nonumber\\
&&+\sum_{\sigma}\bar{d}_{\sigma}(t)(i\partial_{t}-\varepsilon_{0})d_{\sigma}(t)+\sum_{\sigma}\{\bar{\xi}_{\sigma}(t)d_{\sigma}(t)+c.c\}]\nonumber\\
&&\end{eqnarray}
 where $\sigma=\pm1$ and we have suppressed the propagation direction indices of the helical edge states for
 brevity. $\{\xi_{\sigma},\bar{\xi}_{\sigma}\}$ are external source fields introduced to generate $G_{dd,\sigma}(t,t')$.
 Integrating out the fermionic fields $\{c_{2k,\sigma},\bar{c}_{2k,\sigma}\}$ and $\{d_{\sigma},\bar{d}_{\sigma}\}$
 we obtain\begin{eqnarray}
S_{eff}&=&\int_{C}dt[\int
dx\sum_{\sigma}\{\bar{\psi}_{1,\sigma}(x,t)i(\partial_{t}+\sigma
v_{F}\partial_{x})\psi_{1,\sigma}(x,t)\}\nonumber\\
&&-\frac{g}{2}\int dx(\sum_{\sigma}\bar{\psi}_{1,\sigma}(x,t)\psi_{1,\sigma}(x,t))^{2}]\nonumber\\
&&-\sum_{\sigma}\int_{C}dt_{1}dt_{2}[\bar{\xi}_{\sigma}(t_{1})+T_{1}\bar{\psi}_{1,\sigma}(0,t_{1})]g_{D\sigma}(t_{1},t_{2})\times\nonumber\\
&&[\xi_{\sigma}(t_{2})+T_{1}^*\psi_{1,\sigma}(0,t_{2})]
\label{eq:Effective action with source}\end{eqnarray}
 which is an effective action for lead-1 electrons only. $g_{D\sigma}(t,t')$ is the Green's function of the
 quantum dot evaluated under the ``dot+lead 2'' action with
 \begin{eqnarray}
 \label{gdsigma}
 g^{R/A}_{D\sigma}(\omega) & = &\frac{1}{\omega-\varepsilon_{0}\pm i\Gamma_2}   \\ \nonumber
 g^<_{D\sigma}(\omega) & = &\frac{2
 i\Gamma_{2}f(\omega-\mu_{2})}{(\omega-\varepsilon_{0})^{2}+\Gamma_2^2},
 \end{eqnarray}
 where $\Gamma_{1(2)}=\pi\rho_{1(2)}|T_{1(2)}|^2$ is the tunnelling width from the dot to lead $1(2)$.
 $G_{dd,\sigma}(t,t')$ can be obtained by taking functional derivatives of the generating
 functional with respect to the $\{\xi_{\sigma},\bar{\xi_{\sigma}}\}$ fields; we obtain
\begin{eqnarray}
 G_{dd,\sigma}(t,t')&=&g_{D\sigma}(t,t')+|T_{1}|^{2}\int_{C}dt_{1}dt_{2}g_{D\sigma}(t,t_{1})\times\nonumber\\
 &&g_{1\sigma eff}(t_{1},t_{2})g_{D\sigma}(t_{2},t')
 \label{eq:reexpress onsite Greens function}
 \end{eqnarray}
 where
 \[ g_{1\sigma eff}(t,t')\equiv-i\left\langle
 T_{C}\hat{\psi}_{1,\sigma}(0,t)\hat{\psi}^{+}_{1,\sigma}(0,t')\right\rangle_{eff}
 \]
 is the Green's function of lead-1 electrons evaluated at $x=0$ according to the effective action
 (\ref{eq:Effective action with source}) in the absence of the external source terms. Different Keldysh
 components of $G_{dd,\sigma}$ can be extracted from \eqref{eq:reexpress onsite Greens function} using Langreth's sum
 rules \cite{Langreth sum rule}. Combining Eqs.\ (\ref{gdsigma}) and \eqref{eq:reexpress onsite Greens function}, we
 obtain for the tunnelling current \eqref{eq:generalized Wingreen
 formula},
 \begin{eqnarray}
 I&=&-e|T_{1}|^{2}|T_{2}|^{2}\sum_{\sigma}\int d\omega\rho_{2}\frac{1}{(\omega-\varepsilon_{0})^{2}+\Gamma_2^2}\times\nonumber\\
 &&\{\mathrm{Im}g_{1\sigma eff}^{<}(\omega)+2f(\omega-\mu_{2})\mathrm{Im}g_{1\sigma eff}^{R}(\omega)\}.
 \label{eq:intermediate current}\end{eqnarray}
 where the only unknown is the effective green's function $g_{1\sigma eff}(t,t')$.


  To evaluate $g_{1\sigma eff}(t,t')$ we express it in the form of a standard Dyson's equation
 \begin{eqnarray}
 g_{1\sigma eff}(t,t')&=&g_{1\sigma}(t,t')+\int_{C}dt_{1}dt_{2}g_{1\sigma}(t,t_{1})\Sigma_{\sigma,T}(t_{1},t_{2})\nonumber\\
 &&\times g_{1\sigma eff}(t_{2},t')
 \label{eq:Dyson's equation}\end{eqnarray}
 where $g_{1\sigma}(t,t')$ is the Green's function of lead-1 electrons at $x=0$ evaluated at $T_1=0$, and is the standard
 $x=0$ Green's function of spinless Luttinger liquid with interaction strength $K$\cite{Giamarchi}. The self-energy term
 in \eqref{eq:Dyson's equation} represents correction coming from the tunnelling term (last term in Eq.(5)). The tunnelling
 current\ (\ref{eq:intermediate current}) can be expressed in terms of the Fourier-transformed self-energy
 $\Sigma_{\sigma,T}(\omega)$ as

\begin{eqnarray}
 I&=&-e|T_{1}|^{2}|T_{2}|^{2}\sum_{\sigma}\int d\omega\rho_{2}\times\nonumber\\
 &&\mathrm{Im}\{\frac{2(f(\omega-\mu_{2})-f(\omega-\mu_{1}))
 g_{1\sigma}^{R}(\omega)-\left|g_{1\sigma}^{R}(\omega)\right|^{2}(\Sigma_{\sigma,T}^{<}(\omega)+2f(\omega-\mu_{2})
 \Sigma_{\sigma,T}^{R}(\omega))}{\left|1-g_{1\sigma}^{R}(\omega)\Sigma_{\sigma,T}^{R}(\omega)\right|^{2}(
 (\omega-\varepsilon_{0})^{2}+\Gamma_2^2)}\}.\label{eq:final current}\end{eqnarray}

 We observe that the tunnelling current is completely determined by the self-energy function $\Sigma_{\sigma,T}(\omega)$.
 In the absence of electron-electron interaction $\Sigma_{\sigma,T}(\omega)=\Sigma^{(0)}_{\sigma,T}(\omega)=
 |T_1|^2g_{D\sigma}(\omega)$ and our main job here is to understand how $\Sigma^{(0)}_{\sigma,T}(\omega)$ is renormalized by
 the electron-electron interaction.


 In the following we shall study $\Sigma_{\sigma,T}$ in the small bias, low temperature limit $T,|\mu_1-\mu_2|<<
 \Gamma_{1(2)}$ using a standard
 bosonization-RG analysis\cite{Kane&Fisher}. In this limit we may keep only the long-time behavior of $g_{D\sigma}(t)$ in
 Eq.\ (\ref{eq:Effective action with source}) and forget about the more complicated intermediate time behaviors, i.e.
 we approximate
 \begin{eqnarray}
  \label{gda}
   g_{D\sigma}^{R(A)}(t) & = &
   e^{i\varepsilon_0t-\Gamma_2|t|}|_{t\rightarrow\infty}\rightarrow0,
   \\ \nonumber
  g_{D\sigma}^<(t) & \sim & 2i{\Gamma_2\over\varepsilon_0^2+\Gamma_2^2}{e^{i\mu_2t}\over t}.
  \end{eqnarray}

 To proceed further we compare the present problem with the problem of directly tunnelling between a Luttinger liquid and a
 Fermi liquid through a simple tunnelling junction barrier $t$. The Fermi liquid fields can be integrated out as what we have
 done to derive Eq.\ (\ref{eq:Effective action with source}). The only difference in the direct tunnelling problem is that
 $g_{D\sigma}(t)$ in Eq.\ (\ref{eq:Effective action with source}) is replaced by
 $g_{0\sigma}^{R(A)}(t)\sim \mp i\pi N(0)\delta(t)$ and $g_{0\sigma}^<(t)\sim2iN(0)e^{i\mu_2t}/t$ where $N(0)$ is the density
 of states on the Fermi surface. Comparing with Eq.\ (\ref{gda}) We see that
 the resonant tunnelling problem reduces to the direct tunnelling problem in this limit if we replace the dimensionless
 tunnelling parameter $tN(0)\rightarrow T_1\sqrt{\Gamma_2\over\varepsilon_0^2+\Gamma_2^2}$ and the scaling behavior of the
 self-energy in the present problem can be inferred from the corresponding
 direct tunnelling problem which has been analyzed using well-developed bosonization-RG technique. We
 obtain immediately
 \[
 T_1(E)\sim (E/E_F)^{1-K_{eff}},  \]
  in this approximation, where $K_{eff}=2K/(1+K)$\cite{Furusaki}. In particular for
 repulsive interaction ($K<1$), $T_1(E\rightarrow0)\rightarrow0$ and the self-energy scales to zero in the infrared regime,
 which makes perturbative RG applicable. In this case it is sufficient to approximate
 $\Sigma_{\sigma,T}\sim\Sigma^{(0)}_{\sigma,T}$ with $T_1\rightarrow T_1(E)$. Physically, the vanishing of self-energy
 is an alternative way to express the well known result that the tunnelling between the Luttinger liquid lead and the Fermi
 liquid lead vanishes in the infrared limit\cite{Kane&Fisher, Furusaki}. We emphasize here that the renormalization of $T_1$
 is restricted only to the self-energy term and does not appear in other places in the current expression. With this approximation we
 obtain
 \begin{widetext}
 \begin{equation}
 I\sim-2e{|T_{1}|^{2}\Gamma_2\over\pi}\sum_{\sigma}\int d\omega(f(\omega-\mu_{2})-f(\omega-\mu_{1}))\frac{
 Im[g_{1\sigma}^{R}(\omega)]}{\left|\omega-\varepsilon_0+i\Gamma_2-|T_1(E)|^2g_{1\sigma}^{R}(\omega)\right|^{2}}.
 \label{eq:final current2}
 \end{equation}
 \end{widetext}
  where $E=max\{T,|\omega-\mu_1|\}$. Eq.\ (\ref{eq:final current2}) is expected to be reliable in both the low temperature,
  small bias limit $(T,|\mu_1-\mu_2|)<<\Gamma_{1(2)}$ and in the high temperature, large bias limit $(T,|\mu_1-\mu_2|)\sim E_F$
  where $T_1(E)\rightarrow T_1$ and the renormalization of $T_1$ becomes unimportant. We shall analyze the current using this
  approximate formula in the following.

\begin{figure}[!htb]
\includegraphics[scale=0.5]{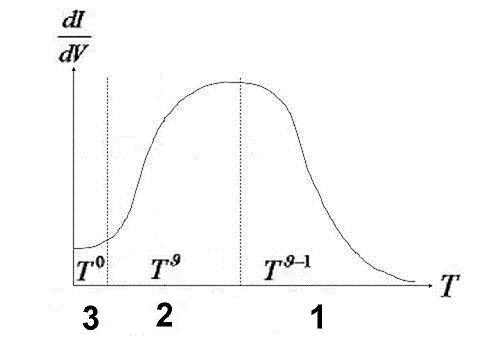}\caption{\label{fig:temperature dependence}The leading-order temperature dependence of differential conductance in different regimes. Regime 1: $T>>\Gamma_{1,2}$; Regime 2: $|eV|<<T<<\Gamma_{1,2}$; Regime 3: $T<<|eV|$.}
\end{figure}

 To study the tunnelling characteristics, we define $eV\equiv\mu_{2}-\mu_{1}$ and
 $\varepsilon\equiv\varepsilon_{0}-\mu_{1}$ and look at the differential conductance $\frac{dI}{dV}$
 as function of $V$ at different temperature regimes. Using the result that $g_{1\sigma}^R(\omega)\sim E^{\theta}$ and
 approximating the Lorentzian function by a $\delta$-function, we find that at very high-temperature regime ($T\gg
 \Gamma_{1,2},\varepsilon_0$), $T_1(E)\rightarrow T_1$ and $\frac{dI}{dV}$ scales with temperature as $T^{\vartheta-1}$,
 in agreement with Furusaki's result \cite{Furusaki}. On the other hand, the low temperature regime $T\ll \Gamma_{1,2}$ can
 be sub-divided into two regions: when $eV\ll T$, the linear differential conductance scales
 as $T^{\theta}(1+aT^{\gamma})$, where $\gamma=\mathrm{min}\{1,\vartheta+2(1-K_{eff})\}$ and $a$ is a temperature-independent
 constant; but when $eV\gg T\rightarrow0$, the current-voltage relation becomes non-linear and ${dI\over dV}$ becomes
 $V-$dependent(see discussion below) at small $V$. The different scaling behaviors are summarized in
 Fig.\ref{fig:temperature dependence}.

 Next we consider (non-linear) differential conductance at zero temperature. First, we observe that in the off-resonance
 region ($|eV|,\Gamma_{1,2}\ll\varepsilon$), the differential conductance goes as $|V|^{\vartheta}(1+a'|V|^{\gamma})$, where
 $a'$ (different from $a$) is a temperature-independent constant. The first term ($|V|^{\theta}$) is in agreement with previous
 result\cite{Chamon & Wen} and the second term $|V|^{\theta+\gamma}$ is the leading correction coming from scaling of $T_1$. Our
 RG amalysis suggests that the leading order ($|V|^{\theta}$ and $T^{\theta}$) scalings are exact as long as
 $g_{1\sigma}^{R}(\omega)\Sigma_{\sigma,T}^{R}(\omega)|_{\omega\rightarrow0}\rightarrow0$ which is the case for repulsive
 electron-electron interaction and is independent of the detailed structure of $\Sigma_{\sigma,T}^{R}(\omega)$. The self-energy
 gives rise only to higher-order corrections to $V/T$ scalings at low temperature and small voltage bias.

 We also observe a resonance peak of differential conductance located around $eV\sim\varepsilon$
 and the differential conductance at $V=0$ is always zero as long as lead 1 is interacting. This is a character of Luttinger
 liquids which is very different for non-interacting electrons ($K=1$), where $\frac{dI}{dV}|_{V\rightarrow0}$ approaches a
 constant. Finally, when $V$ is far away from the resonant point, the differential conductance falls as
 $|V|^{\vartheta-2\times\mathrm{max}\{1,\vartheta+2(1-K_{eff})\}}$.

\begin{figure}
\includegraphics[scale=0.3]{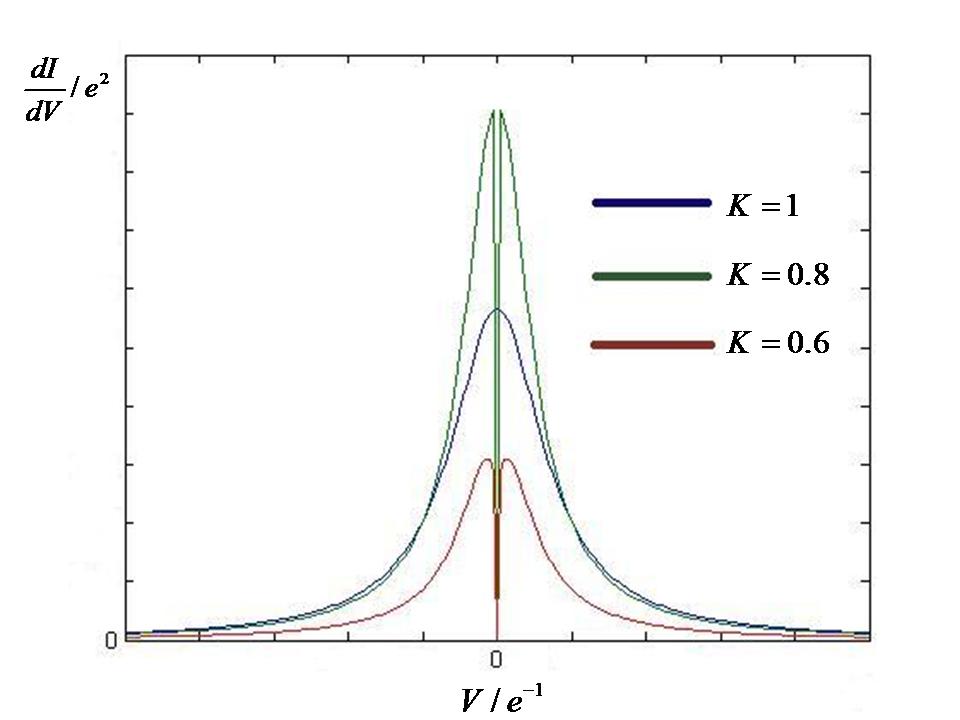}\caption{\label{fig:zero temperature}(color online) Differential conductance at zero temperature with $\varepsilon=0$}
\end{figure}

 To conclude, we have extended Meir-Wingreen's current formula to the case with one interacting lead in this paper, and have
 used it to calculate the tunnelling current from the edge of a 2D topological insulator through a quantum dot to a normal
 Fermi liquid lead. The formulation allows us to construct a current expression in terms of the self-energy of a local
 Green's function. The differential conductance in different temperature regimes is analyzed using perturbative RG
 in this paper, which is believed to be reliable when both temperature and voltage bias are much smaller than $\Gamma_{1(2)}$
 and if the electron-electron interaction in lead 1 is repulsive. Based on the RG result, an approximate formula for the
 current qualitatively valid over whole temperature/volatge range is proposed. The formula reproduces results of earlier works at high
 temperature and produces exact results at low temperature for repulsive electron-electron interaction and small bias. For
 attractive interaction the scaling breaks down at low enough energy $E$ suggesting that qualitatively new $dI/dV$ behavior
 is expected at low energy. Our approach offers a new theoretical tool of analysing (non-equilibrium) DC transports which
 can be extended to other systems with both Luttinger and Fermi liquid leads.

 We thank K. T. Law, Zhengxin Liu and C. Chan for useful discussions. This work is supported by HKRGC grant
 HKUST3/CRF/09.


\begin{thebibliography}{23}






\bibitem{topological insulator1} C. L. Kane and E. J. Male, Science,
\textbf{314}, 1692 (2006).

\bibitem{topological insulator2} C. L. Kane, Nature, \textbf{4},
348 (2008).

\bibitem{SCZhang} C. Wu, B. A. Bernevig and S. C. Zhang, Phys. Rev. Lett. \textbf{96}, 106401 (2006).

\bibitem{Kane&Fisher} C. L. Kane and M. P. A. Fisher, Phys, Rev. B \textbf{46}, 15233 (1992).

\bibitem{Furusaki&Nagaosa} A. Furusaki and N. Nagaosa, Phys. Rev. B \textbf{47}, 3827 (1993).

\bibitem{Fabrizio&Gogolin} M. Fabrizio and A.O. Gogolin, Phys. Rev. B \textbf{50}, 17732 (1994).

\bibitem{Res-Tun in FQH} J. A. Simmons, H. P. Wei, L. W. Engel, D.
C. Tsui and M. Shayegan, Phys. Rev. Lett. \textbf{63}, 1731 (1989).

\bibitem{Chamon & Wen} C. de C. Chamon and X. G. Wen, Phys. Rev.
Lett. \textbf{70}, 2605 (1993).

\bibitem{Furusaki} A. Furusaki, Phys. Rev. B \textbf{57}, 7141 (1998).

\bibitem{Luttinger Liquid} J. M. Luttinger, J. Math. Phys. N. Y.
\textbf{4}, 1154 (1963).

\bibitem{vic} K.T. Law, C.Y. Seng, P.A. Lee and T.K. Ng, \prb {\bf 81}, 041305 (2010)

\bibitem{Kamenev} A. Kamenev and A. Levchenko, Adv. Phy. \textbf{58},
197 (2009).

\bibitem{Giamarchi} See, e.g., T. Giamarchi, \textit{Quantum Physics
in One Dimension }(Oxford University Press, Cambridge, 1998).

\bibitem{My thesis} C. Y. Seng, Mphil Thesis, The Hong Kong University of Science and
Technology (2010).

\bibitem{Meir&Wingreen&Jauho} A. P. Jauho, N. S. Wingreen and Y.
Meir, Phys. Rev. B \textbf{50}, 5528 (1994).

\bibitem{Vurkevich} I. V. Lerner and I. V. Vurkevich, in
\textit{Nanophysics: Coherence and Transport}, edited by H.
Bouchiat, Y. Gefen, S. Gueron, G. Montambaux and J. Dalibard
(Elsevier, New York, 2005)

\bibitem{Langreth sum rule} D. C. Langreth, in \textit{Linear and
Nonlinear Electron Transport in Solids}, Vol.17 of\textit{ Nato
Advanced Study Institute, Series B: Physics}, edited by J. T.
Devreese and V. E. Van Doren (Plenum, New York, 1976).

\bibitem{Wen's book} X. G. Wen, \textit{Quantum Field Theory of Many-Body
Systems} (Oxford University Press, 2004).






\end{thebibliography}
\end{document}